\documentclass[runningheads]{llncs}
\usepackage[T1]{fontenc}
\usepackage{graphicx}

\usepackage[show]{chato-notes}
\usepackage{booktabs}
\usepackage{hyperref}

\usepackage{graphicx}
\usepackage[numbers,sort&compress]{natbib}
\usepackage{balance}
\usepackage{nicefrac}
\usepackage{multirow}
\usepackage{xspace}
\usepackage{booktabs}
\usepackage{makecell}
\usepackage{moresize}
\usepackage{tabularx}
\usepackage{adjustbox}
\usepackage[normalem]{ulem}
\usepackage{siunitx}
\usepackage{tikz}
\usepackage{amsmath,amsfonts}
\usepackage{tablefootnote}
\usepackage{svg}
\usepackage{multirow}

\newcommand{\colbert}{\textsf{ColBERT}}
\newcommand{\constbert}{\textsf{ConstBERT}}
\newcommand{\esplade}{\textsf{ESPLADE}}
\newcommand{\retromae}{\textsf{RetroMAE}}

\newcommand{\crc}[1]{#1}

\newcommand{\dev}{\textsf{Dev}\xspace}
\newcommand{\trecdl}{\textsf{TREC 2019}\xspace}
\newcommand{\trecdltw}{\textsf{TREC 2020}\xspace}

\newcommand{\bfx}{\fontseries{b}\selectfont}

\begin{document}
\title{Efficient Constant-Space Multi-Vector Retrieval}

\author{Sean MacAvaney\inst{1} \and Antonio Mallia\inst{2} \and Nicola Tonellotto\inst{3}}
\authorrunning{MacAvaney \and Mallia \and Tonellotto}
\institute{University of Glasgow, UK \and
Pinecone, US
\and
University of Pisa, Italy}
\maketitle              %
\begin{abstract}

Multi-vector retrieval methods, exemplified by the ColBERT architecture, have shown substantial promise for retrieval by providing strong trade-offs in terms of retrieval latency and effectiveness. However, they come at a high cost in terms of storage since a (potentially compressed) vector needs to be stored for every token in the input collection. To overcome this issue, we propose encoding documents to a fixed number of vectors, which are no longer necessarily tied to the input tokens. Beyond reducing the storage costs, our approach has the advantage that document representations become of a fixed size on disk, allowing for better OS paging management. Through experiments using the MSMARCO passage corpus and BEIR with the ColBERT-v2 architecture, a representative multi-vector ranking model architecture, we find that passages can be effectively encoded into a fixed number of vectors while retaining most of the original effectiveness.%


\vspace{0.6em}
\hspace{5em}\includegraphics[width=1.25em,height=1.25em]{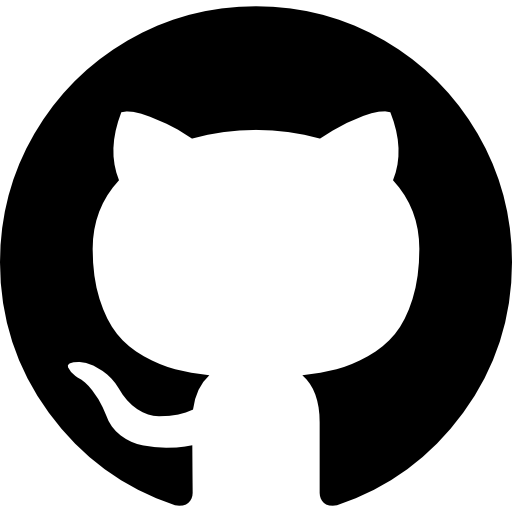}\hspace{.3em}
\parbox[c]{\columnwidth}
{
    \vspace{-.55em}
    \href{https://github.com/pisa-engine/ConstBERT}{\nolinkurl{https://github.com/pisa-engine/ConstBERT}}
}
\vspace{-1.2em}
\end{abstract}
\section{Introduction}

Pre-trained contextualized language models, such as BERT~\cite{bert}, learn semantic embeddings from word contexts, enabling them to better capture the relevance of documents with respect to the queries. Notably, they outperform classical ranking approaches~\cite{lin2020pretrained}. More specifically, cross-encoders concatenate a query and a document texts, and feed them into BERT to compute the query document similarity scores, while bi-encoders compute compact representations of documents as real-valued vectors, both for queries and documents, and the query-document similarity is computed using the cosine similarity or the inner product between query and document embeddings.

Cross-encoder can be computationally expensive for estimating query-document similarities due to the complexity of the underlying transformer neural network~\cite{khattab2020colbert,Hofsttter2019LetsMR,snrm,hofstaetter2020_crossarchitecture_kd}. On the other side, bi-encoders are much more efficient from a computational perspective, since all document embeddings can be pre-computed offline and store in specialised vector indexes such as FAISS~\cite{JDH17}. Instead of relying on a single vector per text as done by bi-encoders, multi-representation systems such as~\cite{khattab2020colbert} use a vector per token in a text, being able to capture more semantics than a single embedding. While ColBERT achieves more effective results than single representations, it comes at the cost of higher response times and memory usage~\cite{iir2021}. 

ColBERTv2~\cite{DBLP:journals/corr/abs-2112-01488} employs a compression method that leverages centroids to represent passage embeddings more efficiently. This method records the ID of the nearest centroid for each embedding and compresses the residuals—the differences between the original embeddings and the centroids—using. This compression strategy helps reduce the storage demands of multi-vector embeddings, but makes retrieval significantly less efficient.
To speed up the search latency of ColBERTv2, PLAID~\cite{santhanam2022plaid} uses a centroid interaction mechanism and centroid pruning to eliminate low-scoring passages early in the search process, thus reducing response times significantly. This approach allows multi-vector retrieval models to maintain retrieval quality while reducing retrieval latency.

XTR (ConteXtualized Token Retriever) \cite{lee2024rethinking} introduces a streamlined approach to multi-vector retrieval by emphasizing efficient token selection during retrieval. Unlike ColBERT’s three-stage process (token retrieval, gathering, scoring), XTR simplifies the retrieval pipeline by training the model to prioritize key document tokens, thus only scoring based on these retrieved tokens. 
Another recent approach worth mentioning is Static Pruning for Multi-Representation Dense Retrieval \cite{acquavia2023static}, which addresses the storage challenges in multi-vector models like ColBERT by pruning embeddings for less impactful tokens. By adapting static pruning techniques traditionally used in sparse indexes to embedding-based indexes, this method reduces storage requirements while maintaining retrieval effectiveness
In addition to pruning techniques, a recent advancement named Token Pooling~\cite{clavie2024reducing} aims to reduce the storage requirements of multi-vector retrieval models like ColBERT. This approach clusters and pools similar token embeddings at indexing time, substantially lowering vector counts without model modifications or query-time processing. 

Multi-vector models like ColBERT are also highly effective as reranking techniques~\cite{macavaney2024reproducibility}, where they refine the ranking of a candidate set generated by simpler retrieval methods, such as BM25. In this setting, ColBERT’s pre-computed, token-level document embeddings allow for efficient late interaction with the query, balancing computational efficiency with strong retrieval performance. This approach leverages the richness of multi-vector representations while maintaining low latency, as document vectors can be cached across queries.

\crc{Recently, MUVERA (Multi-Vector Retrieval Algorithm) \cite{dhulipala2024muvera} introduced a mechanism to bridge the gap between single-vector and multi-vector retrieval. MUVERA employs Fixed Dimensional Encodings (FDEs) to approximate multi-vector similarities, enabling the use of optimized Maximum Inner Product Search (MIPS) solvers. This approach significantly enhances retrieval efficiency compared to methods like PLAID. Although MUVERA achieves a good approximation of PLAID using a single-vector representation and MIPS operations, which makes its retrieval algorithm faster, it is able to do so only by employing large high-dimensional vectors. This results in substantial memory consumption, presenting a trade-off between retrieval speed and storage efficiency.}

In this paper, we propose a novel approach\crc{, \constbert{},} to reduce the storage footprint of multi-vector retrieval by encoding each document with a fixed, smaller set of learned embeddings. Instead of relying on token-level embeddings across all document tokens, we introduce a pooling mechanism that projects these token embeddings into a reduced set of document-level embeddings, each capturing distinct semantic facets.
This learned pooling reduces the number of embeddings stored per document, achieving considerable space savings in the index while retaining retrieval effectiveness. 
The fixed number of vectors per document also eases the use of \crc{\constbert{}} as a reranking method, simplifying integration with initial retrieval systems and allowing efficient late interaction with pre-computed document representations. This reduction is complementary to other methods, such as dimensionality reduction, and enables efficient memory alignment with OS-level paging, ultimately improving both storage efficiency and query processing speed.

\section{\crc{\constbert{}:} Multi-Vector Compression}

Given a query $q$, our task is to retrieve relevant documents $d$ from a corpus $D$ by ranking them with a relevance scoring function $s(q,d)$. Queries and documents are sequences of tokens
from a given vocabulary. Each document $d$ comprises $M$ tokens, and each query $q$ comprises $N$ tokens, with padding/masking tokens if necessary.
In a multi-representation dense IR system, any token is represented by a $k$-dimensional real-valued vector, called embedding. Let $q_1, \ldots, q_N$ denote the token embeddings for the query $q$, and $d_1, \ldots, d_M$ denote the token embeddings for the document $d$. The relevance score $s(q,d)$ between the query $q$ and the document $d$ is computed with a late interaction mechanism:
\begin{align*}
    s(q,d) = \sum_{i=1}^{N}\max_{j=1,\ldots,M}q_i^Td_j.
\end{align*}

This late interaction mechanism sums up the contributions of the most relevant document token for each
query token. For each query token, the $\max$ operator can be interpreted as an heuristic pooling mechanism over the token embeddings of the document. Instead of relying on this heuristic pooling across all document tokens, in this paper we propose a new \textit{learned} pooling, where instead of using the document embeddings $d_1,\ldots,d_M$, i.e., an embedding per document token, we use $C < M$ new embeddings $\delta_1, \ldots, \delta_C$, that are learned with an additional projection layer with parameters $W \in \mathbb{R}^{Mk \times Ck}$:
\begin{align*}
    [\delta_1 | \cdots | \delta_C] = W^T [d_1 | \cdots | d_M].
\end{align*}

This layer \crc{is learned end-to-end during training} takes as input the token embeddings of a document computed by the multi-representation dense IR system, and with a linear transformation projects them in a fixed number of embeddings, of the same dimensions. In doing so, the new embeddings can be seen as different single document-level embeddings, each one encoding some semantic facet of the document, given its token embeddings.
The relevance score $s(q,d)$ between the query $q$ and the document $d$ is computed now as:
\begin{align*}
    s(q,d) = \sum_{i=1}^{N}\max_{j=1,\ldots,C}q_i^T\delta_j.
\end{align*}
As a result, the total number of embeddings per document to store in the embedding index decreases by a factor $M/C$. This reduces the space required to store the index on disk and in main memory, as well as the query processing time. This space reduction is orthogonal to any further space reduction obtained, for example, by reducing the number of dimensions $k$ per embedding. Both reductions can be further exploited to align the space occupancy per document to the memory page size, so to exploit more efficiently the underlying memory management mechansisms provided by the operating system.

\section{Experimental Results}
In this section, we evaluate the performance and efficiency of our proposed fixed-vector model, denoted as \crc{\constbert$_C$, where $C$} represents the number of fixed embeddings per document. We compare it against the baseline, \colbert, which uses token-level embeddings for each document token. 
\subsection{Experimental Setup}

    \paragraph{Datasets \& Queries.}
    Our experimental framework utilizes the MSMARCO~v1 passage corpus \cite{Bajaj2016Msmarco}, which consists of approximately 8.8 million passages. To assess both the effectiveness and efficiency of query processing, we benchmark our approach against established methodologies using the MSMARCO Dev Queries, as well as datasets from the TREC Deep Learning Tracks of 2019 and 2020 \cite{trec19dloverview, trec20dloverview}. Furthermore, we conduct evaluations on an additional 13 collections drawn from the BEIR benchmark \cite{thakur2021beir}, allowing for a comprehensive analysis across datasets.
    \paragraph{Metrics.}
    To evaluate effectiveness, we employ the official metrics designated for each query set: Mean Reciprocal Rank at cutoff 10 (MRR@10) for MSMARCO Dev queries and Normalized Discounted Cumulative Gain at cutoff 10 (NDCG@10) for both TREC queries and the BEIR benchmark. Additionally, we report recall across varying cutoff thresholds for the MSMARCO experiments. For efficiency analysis, we compute the Mean Response Time (MRT) for both the Dev and TREC queries, measured in milliseconds. Additionally, we examine the index sizes to demonstrate the substantial storage efficiency gains achieved by our method.

    \paragraph{Implementations.}
    \crc{\constbert{}} has been trained following the approach proposed by \citet{DBLP:journals/corr/abs-2112-01488}.
    \colbert{$_{SP}$} refers to the modification of \colbert{} proposed by \citet{acquavia2023static}, where token embeddings  are statically pruned at indexing time.
    \retromae{}~\cite{xiao2022retromae} is a single-representation dense retrieval model for which we have used the official checkpoint\footnote{https://huggingface.co/Shitao/RetroMAE\_MSMARCO\_distill}.

    \paragraph{Platform.}
All experiments were carried out in memory on a Linux system, using a single processing thread. The hardware configuration included dual 2.8 GHz Intel Xeon CPUs and 256 GiB of RAM. For end-to-end retrieval experiments, we used the official PLAID~\cite{santhanam2022plaid, macavaney2024reproducibility} codebase\footnote{https://github.com/stanford-futuredata/ColBERT}. In our two-stage retrieval experiments, we used BMP~\cite{mallia2024faster} and \textsf{efficient SPLADE}~\cite{lassance2022efficiency} for candidate generation and then we performed reranking using our \crc{\constbert$_{32}$} model. We tested other learned sparse models as first-stage retrieval methods, such as DeeperImpact~\cite{deepimpact,deeperimpact}, and obtained similar results; however, we did not include them in the main results section due to limited space.
Our code is written in Python and is available at \url{https://github.com/pisa-engine/ConstBERT}.

\subsection{Overall}
Table \ref{tab:performance_metrics} presents the results on MSMARCO with different configurations of \crc{\constbert$_C$ (varying the number of fixed embeddings per document $C$)} and the baseline \colbert. As expected, \constbert’s performance improves with larger token-level configurations, but at the cost of substantial increases in index size. Our proposed \constbert$_{32}$ model achieves comparable MRR on the development set and NDCG@10 on both TREC 2019 and TREC 2020 benchmarks, while using a fixed number of vectors per document, which allows for more efficient storage. \crc{\constbert$_{C}$ has a variety of tradeoffs compared to the existing static pruning approach \colbert$_{SP}$. The performance of \colbert$_{SP}$ lines on the storage-effectiveness Pareto frontier set by various settings of $C$. On the one hand, \constbert$_{C}$ offers advantages in flexibility since it can be tuned directly to a target (and constant-space) representation. On the other hand, it requires re-training learn the weights $W$, while \colbert$_{SP}$ does not require retraining.}

To evaluate the robustness of \crc{\constbert$_C$} across different retrieval tasks, we further assess it on the BEIR benchmark (Table \ref{tab:beir}). The results highlight that our model performs competitively with \colbert on most tasks, achieving comparable or even superior NDCG@10 scores while requiring much less storage. 

A major advantage of our fixed-vector approach lies in its reduced storage footprint. Unlike \colbert, which scales linearly with the number of token embeddings per document, \crc{\constbert$_C$} maintains a consistent index size by using a fixed number of embeddings. This efficiency extends across the BEIR datasets, with our approach consistently reducing index sizes by over 50\% compared to \colbert at equivalent effectiveness.

The reduced storage requirement of \crc{\constbert$_C$} also translates into lower memory usage and faster retrieval times, as fewer embeddings need to be processed per query-document pair. This efficiency gain is particularly advantageous when using \crc{\constbert$_C$} as a reranking method, where computational speed is crucial.

\begin{table}[t]
\caption{Effectiveness metrics and index space consumption on different query sets for the MSMARCO benchmark. }
\label{tab:performance_metrics}
\centering
\resizebox{\textwidth}{!}{%
\begin{tabular}{lcccccccccccccc}
\toprule
& \multicolumn{4}{c}{\dev}  & \multicolumn{4}{c}{\trecdl} & \multicolumn{4}{c}{\trecdltw} \\
\midrule
   & \multirowcell{2}{Index Size} & \multirowcell{2}{MRR} & \multicolumn{3}{c}{Recall} & \multirowcell{2}{NDCG@10} & \multicolumn{3}{c}{Recall} & \multirowcell{2}{NDCG@10} & \multicolumn{3}{c}{Recall} \\ 
  \cmidrule(lr){4-6}  \cmidrule(lr){8-10} \cmidrule(lr){12-14} 
  &&  & 50 & 200 & 1000 &  & 50 & 200 & 1000 & & 50 & 200 & 1000\\
\midrule
\colbert  & 22G & 39.99 & 86.52 & 94.47 & 97.34 & 74.64 & 45.64 & 68.88 & 83.11&  73.99 & 53.80 & 72.64& 82.70\\ 
\colbert$_{SP}$ & 14G & 39.12 & 85.81 & 93.80 & 97.00  & 74.42 & 45.41 & 66.58 & 79.74&  72.36 & 52.94 & 71.74 & 82.04 \\
\midrule
\constbert$_{16}$  & 5G &37.84 & 84.04 & 91.74 & 94.11 & 71.15 & 41.38 & 61.31 & 72.62 & 73.75 & 48.98  & 65.80 & 73.89  \\ 
\constbert$_{32}$ & 11G &  39.04 & 85.86 & 93.72 & 96.34 & 73.14 & 44.93 & 65.46 & 78.37 & 73.29 & 51.57 & 69.74 & 79.14 \\ 
\constbert$_{64}$ & 20G & 39.15 & 86.27 & 94.06 & 96.90 & 74.29 & 46.07 & 66.97& 79.64 & 73.47 & 52.85 & 71.98 & 81.62\\ 
\constbert$_{128}$  & 40G & 39.53 & 86.46 & 94.39 & 97.29 & 74.37 & 46.79 & 68.05 & 81.28 & 73.31 & 52.04 & 71.80 & 82.36 \\
\bottomrule
\end{tabular}
}
\end{table}

\begin{table}[h]
\centering
\caption{NDCG@10 and index space consumption on 13~datasets of the BEIR benchmark. }
\label{tab:beir}
\resizebox{\textwidth}{!}{%
\begin{tabular}{lrrrrrrrrrrrrr}
\toprule
      & arguana  & cfever & dbpedia & fever    & fiqa     & hotpot &  nf & nq       & quora    & scidocs  & scifact  & covid & touche \\
\midrule
& \multicolumn{13}{c}{\bf nDCG@10} \\
\midrule
\colbert &   0.452 &\bfx0.163 &\bfx0.434 &\bfx0.751 &\bfx0.338 & \bfx0.679 &\bfx0.329  &\bfx0.554 &\bfx0.846 &    0.154 &\bfx0.638 &    0.705   &\bfx0.261 \\
(\retromae) & 0.366 &\bfx0.197 &\bfx0.469 & 0.719 &\bfx0.339 & 0.602 & 0.311 & 0.526 &\bfx0.864 & 0.139 &\bfx0.647 & 0.649 &\bfx0.326 \\

\constbert$_{32}$       &    0.451 &  0.142 &   0.418 &    0.696 &    0.312 &    0.621 &    0.327  &    0.534 &    0.821 &    0.156 &    0.607 &\bfx0.745   &  0.260 \\
\constbert$_{64}$       &\bfx0.465 &  0.139 &   0.420 &    0.695 &    0.335 &    0.639 &    0.323  &    0.537 &    0.818 &\bfx0.159 &    0.631 &    0.719   &  0.250 \\
\midrule
& \multicolumn{13}{c}{\bf Index Size} \\
\midrule 
\colbert &    49M &     17G &     11G &    17G &    232M &      12G &    23M &    8.3G &\bfx0.3G &    143M &    32M &    770M &     1.5G \\
(\retromae) &  26M &     17G &     15G &    17G &    170M &      15G &    11M &    7.7G &    1.5G &     77M &    16M &    505M &     2.3G \\
\constbert$_{32}$       &\bfx13M &\bfx  6G &\bfx  5G &\bfx 6G &\bfx 72M &\bfx   6G &\bfx 5M &\bfx3.2G &    0.6G &\bfx 33M &\bfx 7M &\bfx212M &\bfx 0.5G \\
\constbert$_{64}$       &    22M &     13G &     11G &    13G &    138M &      12G &    10M &    6.1G &    1.2G &     64M &    15M &    408M &     0.9G \\
\bottomrule
\end{tabular}
}
\end{table}

\begin{table}[h]
\caption{Effectiveness metrics and mean response time (MRT,
in ms) for top-$10$ retrieval using  PLAID vs. two-stage on
\dev{} Queries, \trecdl, and \trecdltw{}.}
\label{tab:rerank}
\centering
\begin{tabular}{llccccc}
\hline
 & \multicolumn{2}{c}{\dev}& \multicolumn{2}{c}{\trecdl{}}   &  \multicolumn{2}{c}{\trecdltw{}}   \\ 
 & MRR & MRT & nDCG@10 & MRT & nDCG@10 & MRT  \\ 
\hline
\colbert & 39.99 & 51.25 & 74.26 & 51.46 & 73.99 & 50.21 \\
\esplade  & 38.75 & 3.07 &  71.33 & 3.13 & 71.14 & 3.20 \\ 
\hspace{0em} + \constbert$_{32}$ & 39.52 & 4.95 & 74.38 & 5.50 & 74.33 & 5.23 \\ %
\hline
\end{tabular}
\end{table}

\subsection{Reranking}
In Table~\ref{tab:rerank}, we evaluate the performance of using \constbert{} as a reranking model instead of employing it in an end-to-end retrieval system. Specifically, we compare PLAID with a two-stage retrieval process incorporating \esplade{} as the model used for candidate generation and our \constbert$_{32}$. The results show the retrieval effectiveness in terms of MRR and nDCG@10, and the average computational efficiency represented by MRT.
Combining \esplade{} with our lightweight $\colbert_{32}$ version balances both performance and efficiency. This combination improves nDCG@10 close to that of the standalone \colbert{} while maintaining MRT below 6 ms, showcasing a practical trade-off.
\constbert$_{32}$ is particularly advantageous not only due to its smaller index size but also because all documents are embedded with the same number of vectors. This uniformity simplifies implementation and allows for optimized memory usage through aligned memory reads, thereby leveraging the underlying memory management mechanisms provided by the operating system more efficiently.

\section{Conclusion}
Our experimental results demonstrate that the fixed-vector approach, \crc{\constbert{}}, effectively balances retrieval effectiveness and storage efficiency. By encoding each document with a fixed, smaller set of learned embeddings, our proposal achieves competitive performance across TREC and BEIR benchmarks, while substantially reducing index size and computational demands. This makes it a scalable and practical solution for real-world information retrieval applications, where both storage efficiency and retrieval speed are essential.

\crc{There are various opportunities for future work in this space. For instance, several studies related to the interpretability of late interaction models have been conducted, given their alignment between tokens and their corresponding representations (e.g.,~\cite{clinchant2,iir2021,DBLP:conf/sigir/WangMTO23}). With our approach, this direct vector-token alignment is no longer present. However, there still may be ways to interpret the interactions, so it may be worth revisiting these studies. Another interesting direction is the application of Pseudo-Relevance Feedback (PRF) with late interaction models (e.g.,~\cite{WANG2022103026}). Future studies could explore whether our approach is complementary to PRF.}

\begin{credits}
\subsubsection{\ackname}
This work was partially supported by
the Spoke ``FutureHPC \& BigData'' of the ICSC – Centro Nazionale di Ricerca in High-Performance Computing, Big Data and Quantum Computing funded by the Italian Government, the FoReLab and CrossLab projects (Departments of Excellence), the NEREO PRIN project (Research Grant no. 2022AEFHAZ) funded by the Italian Ministry of Education and Research (MUR), and the FUN project (SGA 2024FSTPC2PN30) funded by the OpenWebSearch.eu project (GA 101070014).
\end{credits}

\bibliographystyle{plainnat}
\bibliography{reference}
\end{document}